\documentclass[showpacs,aps,graphicx,twocolumn]{revtex4}
\usepackage{amsmath}
\usepackage{mathrsfs}
\usepackage{amsfonts}
\usepackage{amssymb}
\usepackage{graphicx}
\usepackage{subfigure}
\usepackage{eufrak}
\usepackage{multirow}
\usepackage{longtable}
\usepackage{float}
\usepackage{bm}
\usepackage{xcolor}
\usepackage[colorlinks,linkcolor=blue,anchorcolor=blue,citecolor=blue,urlcolor=blue]{hyperref}
\usepackage{soul,color,xcolor}

\begin{document}

\title{Efficient and deterministic high-dimensional controlled-swap gates on hybrid linear optical systems with high fidelity}

\author{Gui-Long Jiang,\textsuperscript{1} Jun-Bin Yuan,\textsuperscript{1} Wen-Qiang Liu,\textsuperscript{2} and Hai-Rui Wei\textsuperscript{1,}}
\email[]{hrwei@ustb.edu.cn}
\address{\textsuperscript{\rm1} School of Mathematics and Physics, University of Science and Technology Beijing, Beijing 100083, China\\
\textsuperscript{\rm2} Center for Quantum Technology Research and Key Laboratory of Advanced Optoelectronic Quantum
Architecture and Measurements (MOE), School of Physics, Beijing Institute of Technology, Beijing 100081, China}

\begin{abstract}

Implementation of quantum logic gates with linear optical elements plays a prominent role in quantum computing due to the relatively easier manipulation and realization.
We present efficient schemes to implement controlled-NOT (CNOT) gate and controlled-swap (Fredkin) gate by solely using linear optics.
We encode the control qubits and target qudits in photonic polarization (two-level) and spatial degrees of freedom ($d$-level), respectively.
Based on the hybrid encoding, CNOT and Fredkin gates are constructed in a deterministic way without any borrowed ancillary photons or measurement-induced nonlinearities.
Remarkably, the number of linear optics required to implement a CNOT gate has been reduced to one polarization beam splitter (PBS), while only $d$ PBSs are necessary to implement a generalized Fredkin gate.
The optical depths of all schemes are reduced to one and dimension-independent.
Besides, the fidelity of our three-qubit Fredkin gate is higher than 99.7\% under realistic conditions, which is higher than the previous schemes.

\end{abstract}

\pacs{03.67.Hk, 03.65.Ud, 03.67.Mn, 03.67.Pp}

\maketitle

\section{Introduction}\label{sec1}

Quantum computing has unparalleled advantages over classical computing in performing information-processing tasks  \cite{1,3,LI,uni-CNOT}, and it has the potential to solve certain discrete computational problems more efficiently; for example, Shor's and Grover's algorithms \cite{Shor,Grover,Grover1,Shor1}.
Universal quantum logic gates are the fundamental building blocks for a quantum computer \cite{1,uni-CNOT}.
Among them, considerable attention has been paid to controlled-NOT (CNOT) gates, controlled-controlled-NOT (Toffoli) gates, and controlled-SWAP (CSWAP, Fredkin) gates. 
This is because arbitrary multiqubit quantum computation can be accurately simulated by CNOT gates and single-qubit rotations \cite{uni-CNOT}, and Fredkin and Toffoli gates form key ingredients of  multiqubit unitary transformation \cite{uni-CNOT}, quantum algorithms \cite{Algorithm}, quantum cryptography \cite{Fingerprinting1,Fingerprinting2}, and quantum fault tolerance \cite{Tolerant}.
Direct optical implementation of Fredkin and Toffoli gates is a more-efficient approach under realistic conditions, instead of synthesis programs \cite{Cost-Fredkin1,Cost-Fredkin2,Cost-Toffoli} that are based on sequences of CNOT and Hadamard gates, leading to Fredkin and Toffoli gates being more susceptible to their environment.

CNOT and Fredkin gates have been experimentally demonstrated in various physical systems \cite{trapped,Fredkin1989,superconductor}.
However, these candidates have certain weaknesses.
For example, superconducting suffers from a short coherence time and weak scalablility, the scalability of ion and high fidelity of neutral atom-based entangling gates is difficult to achieve experimentally, and spin in solids is challenged by inefficiency and impracticality.
Compared with these complicated physical mediums, photonic qubits have several advantages, such as negligible decoherence, diverse qubitlike degrees of freedom (DOFs) \cite{Wei2020,Frequency,multi-qubit,time}, and easy single-qubit operations with linear optical devices \cite{single-qubit}.
Moreover, the lack of interactions between individual photonic qubits has been remedied use of a cross-Kerr-bus \cite{Cross-Kerr1,Cross-Kerr2} or linear-optics networks \cite{KLM,experimental2004,experimental2005,linear2006,linear2008}.
In particular, the architecture based on linear optics is the simplest to achieve with current technology, and linear optics-based schemes may be optimized use of high-dimensional Hilbert spaces \cite{Wei2020} or multiple DOFs of a single photon \cite{Hybrid2015,Hybrid2019}.

In high-dimensional quantum computing, blocks of these basic logic gates need to be arranged in order, which results in quantum circuits that are extremely complex and much low  success probability.
Integrated photonic circuits, based on a Mach-Zehnder interferometer (MZI) \cite{MZI}, might sidestep this obstacle \cite{Integrate1,Integrate2,Integrate3,Integrate4,Integrate5,Integrate6,Review}.
By using  path- or polarization-encoded qubits, such schemes enable large-scale quantum computations by decomposing arbitrary unitary transformations into multiple MZIs \cite{MZI,Integrate3,Decomposition}.
However, the imperfect linear optics and the higher number of paths make the fidelity of the quantum transformation low and make the gates more vulnerable to their environment.
In contrast to constructions based on a single photonic DOF, constructions with multiple DOFs not only increase the capacity of the quantum information carried but also reduce the number of linear optics and transmission losses \cite{Hybrid2015,Hybrid2019}.
Tremendous progress has been made with regard to deterministic quantum transformation with several photonic DOFs. 
Polarization-orbital-angular-momentum Toffoli and Fredkin gates were experimentally demonstrated in 2021 \cite{Single-photon1,Single-photon2}. 
In 2022, Meng \cite{path-polarization1} proposed a Fredkin gate based on polarization-entangled photon pairs with high fidelity. 
Chen \emph{et al}. \cite{path-polarization2} optimized the unitary transformations in terms of a MZI and optical depth by using spatial and polarization DOFs.

In this paper, we propose potentially practical schemes to implement CNOT and controlled-SWAP gates that carry a two-level polarization-based control qubit and a multilevel spatially encoded target qudit.  The schemes are all constructed with the use of solely linear optics.
By conversion of spatial operations to polarization-dependent operations, the CNOT gate $U_{\text{CNOT}}^{2,2}$ can be constructed with only one polarizating beam splitter (PBS), and the controlled-swap gate $U_{\text{CSWAP}}^{2,d,d}$ requires $d$ PBSs; see Fig. \ref{Fig.1} and Fig. \ref{Fig.4}.
Compared wth the scheme shown in Ref. \cite{path-polarization1}, we decrease the number of linear optics required in the construction of the Fredkin gate $U_{\text{CSWAP}}^{2,2,2}$ (CNOT gate $U_{\text{CNOT}}^{2,2}$), from 14 (5) to 2 (1), and we reduce the optical depth from 11 (5) to 1 (1).
Moreover, we also increase the fidelity of $U_{\text{CSWAP}}^{2,2,2}$ to more than $99.7\%$.
The deterministic gate mechanisms, fewer quantum resources (linear optics), higher fidelities, no ancillary photons, and dimension-independent optical depth make our linear-optics architectures feasible with current technology.

\section{Low-cost implementations of controlled-NOT gate and Fredkin gates}\label{sec2}

In this section, we propose two simple constructions of CNOT  and Fredkin gates with only linear optics by suitably encoding in polarization and spatial DOFs. 
The implementation of Fredkin gates is then extended to arbitrary higher-dimensional quantum systems.
%

\subsection{The linear optical controlled-NOT gate}\label{sec2.1}

In the $\{|00\rangle,|01\rangle,|10\rangle,|11\rangle\}$ basis of four-dimensional Hilbert space, the matrix form of the CNOT gate is given by
\begin{eqnarray}             \label{eq1}
U_{\text{CNOT}}^{2,2}=\left(
  \begin{array}{cccc}
    1 & 0 & 0 & 0 \\
    0 & 1 & 0 & 0 \\
    0 & 0 & 0 & 1 \\
    0 & 0 & 1 & 0 \\
  \end{array}
\right).
\end{eqnarray}

\begin{figure} [htbp]
  \centering
  \includegraphics[width=7 cm]{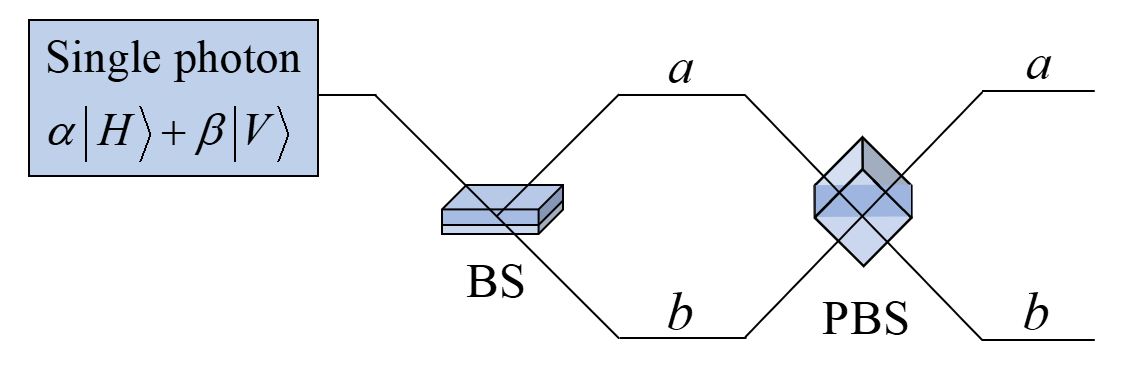}
  \caption{The polarization-spatial controlled-NOT gate with linear optics.
  Single photon $(\alpha|H\rangle+\beta|V\rangle)$ emitted from spatial mode $|a\rangle$.
  The PBS transmits the horizontal component and reflects the vertical component of photons.
  The $\textrm{BS}$ has variable reflectivity.}\label{Fig.1}
\end{figure}

We first prepare an input state of the CNOT gate to demonstrate the gate operation.
As shown in Fig. \ref{Fig.1}, the initial state of the single photon is given by
\begin{eqnarray}\label{eq2}
|\phi_{\text{initial}}\rangle=(\alpha \hat a^{\dag}_{H}+\beta \hat a^{\dag}_{V})|\textrm{vac}\rangle.
\end{eqnarray}
Here and afterward, $\hat m^{\dag}$ denotes the creation operator in spatial mode $m$, $H$ and $V$ refer to horizontal and vertical polarization modes of the photon, respectively, ``vac'' represents the vacuum state, $\alpha$ and $\beta$ are the arbitrary coefficients of the initial state and satisfy $|\alpha|^{2}+|\beta|^{2}=1$.

After the single photon passes through a beam splitter (BS), which results in the transformations
\begin{eqnarray}\label{eq3}
\begin{aligned}
\hat a^{\dag}_{H} \xrightarrow{\textrm{BS}} \gamma \hat a^{\dag}_{H} + \delta \hat b^{\dag}_{H},\;\;
\hat a^{\dag}_{V} \xrightarrow{\textrm{BS}} \gamma \hat a^{\dag}_{V} + \delta \hat b^{\dag}_{V},
\end{aligned}
\end{eqnarray}
where $|\gamma|^{2}+|\delta|^{2}=1$, $|\phi_\text{initial}\rangle$ is transformed into
\begin{eqnarray}\label{eq4}
\begin{aligned}
|\phi_{\text{in}}\rangle=(\alpha \gamma \hat a^{\dag}_{H}+\alpha \delta \hat b^{\dag}_{H}+\beta \gamma \hat a^{\dag}_{V}+\beta \delta \hat b^{\dag}_{V})|\textrm{vac}\rangle.
\end{aligned}
\end{eqnarray}
Then we encode the state of polarizations and spatial modes as
\begin{eqnarray}\label{eq5}
\begin{aligned}
|V\rangle\equiv|0\rangle_{1},|H\rangle\equiv|1\rangle_{1},
|a\rangle\equiv|0\rangle_{2},|b\rangle\equiv|1\rangle_{2}.
\end{aligned}
\end{eqnarray}
%
Thus, we can get a two-qubit input state of the CNOT gate
\begin{eqnarray}\label{eq6}
\begin{aligned}
|\phi_{\textrm{in}}\rangle=\alpha \gamma |10\rangle_{12}+\alpha \delta |11\rangle_{12}+\beta \gamma |00\rangle_{12}+\beta \delta |01\rangle_{12},
\end{aligned}
\end{eqnarray}
which corresponds to the general input state described in Eq. (\ref{eq4}).
Note that our encoding given by Eq. (\ref{eq5}) is the opposite of the encoding $|H\rangle=|0\rangle_{1},|V\rangle=|1\rangle_{1},|b\rangle=|0\rangle_{2},|a\rangle=|1\rangle_{2}$ in Ref. \cite{path-polarization1}.

On the basis of the encoded input states $|\phi_{\textrm{in}}\rangle$, one can find that the operation $U_{\text{CNOT}}^{2,2}$ can be achieved by use of only one PBS; see Fig. \ref{Fig.1}.
Here the PBS transmits the  horizontal component and reflects the vertical component of photons, i.e.,
\begin{eqnarray}\label{eq7}
\begin{aligned}
&\hat a^{\dag}_{H} \xrightarrow{\textrm{PBS}} \hat b^{\dag}_{H},\;\;
\hat a^{\dag}_{V} \xrightarrow{\textrm{PBS}} \hat a^{\dag}_{V},\\
&\hat b^{\dag}_{H} \xrightarrow{\textrm{PBS}} \hat a^{\dag}_{H},\;\;
\hat b^{\dag}_{V} \xrightarrow{\textrm{PBS}} \hat b^{\dag}_{V}.
\end{aligned}
\end{eqnarray}
Thus, after $\hat a^{\dag}_{H}$, $\hat a^{\dag}_{V}$, $\hat b^{\dag}_{H}$, and $\hat b^{\dag}_{V}$ rejoin at the PBS,  the state $|\phi_{\textrm{in}}\rangle$ described in Eq. (\ref{eq4}) becomes
\begin{eqnarray}\label{eq8}
\begin{aligned}
|\phi_{\textrm{out}}\rangle=(\alpha \gamma \hat b^{\dag}_{H}+\alpha \delta \hat a^{\dag}_{H}+\beta \gamma \hat a^{\dag}_{V}+\beta \delta \hat b^{\dag}_{V})|\textrm{vac}\rangle,
\end{aligned}
\end{eqnarray}
which corresponds to the output state
\begin{eqnarray}\label{eq9}
\begin{aligned}
|\phi_{\textrm{out}}\rangle=\alpha \gamma |11\rangle_{12}+\alpha \delta |10\rangle_{12}+\beta \gamma |00\rangle_{12}+\beta \delta |01\rangle_{12}.
\end{aligned}
\end{eqnarray}

\subsection{The linear optical controlled-SWAP gate}\label{sec2.2}

The matrix form of the Fredkin gate is given by
\begin{eqnarray}             \label{eq10}
U_{\text{CSWAP}}^{2,2,2}=\left(
  \begin{array}{cccccccc}
    1 & 0 & 0 & 0 & 0 & 0 & 0 & 0 \\
    0 & 1 & 0 & 0 & 0 & 0 & 0 & 0 \\
    0 & 0 & 1 & 0 & 0 & 0 & 0 & 0 \\
    0 & 0 & 0 & 1 & 0 & 0 & 0 & 0 \\
    0 & 0 & 0 & 0 & 1 & 0 & 0 & 0 \\
    0 & 0 & 0 & 0 & 0 & 0 & 1 & 0 \\
    0 & 0 & 0 & 0 & 0 & 1 & 0 & 0 \\
    0 & 0 & 0 & 0 & 0 & 0 & 0 & 1 \\
  \end{array}
\right),
\end{eqnarray}
in the $\{|000\rangle, |001\rangle, |010\rangle, |011\rangle, |100\rangle, |101\rangle,$
$|110\rangle,|111\rangle\}$ basis.

\begin{figure} [htbp]
  \centering
  \subfigure[State preparation]{
  \includegraphics[width=5 cm]{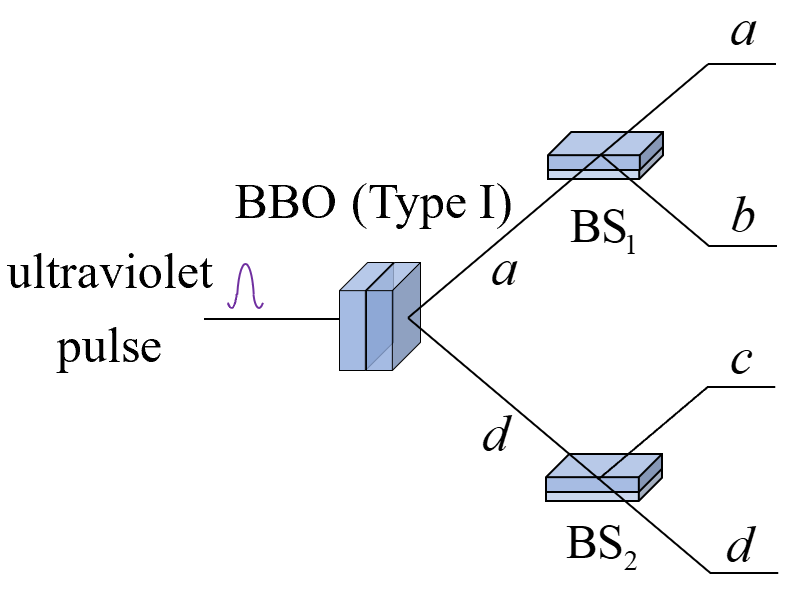}
  \label{2a}}
  \subfigure[Controlled-swap gate]{
  \includegraphics[width=4.2 cm]{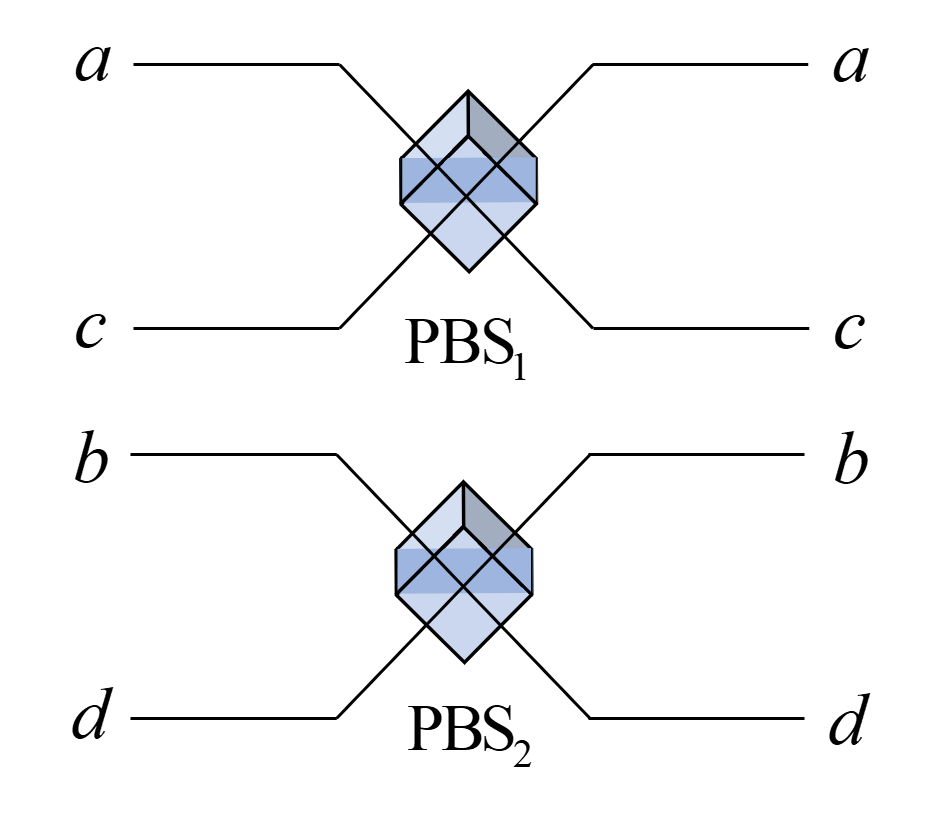}
  \label{2b}}
  \caption{(a) Experimental setup to create the input state of Fredkin gates. (b) The polarization-spatial controlled-SWAP gate with linear optics.}
  \label{Fig.2}
\end{figure}

The schematic setup, shown in Fig. \ref{2a}, is designed to prepare the input normalization state $|\psi_\text{in}\rangle$ of the controlled-SWAP gate. Here
\begin{eqnarray}\label{eq11}
\begin{aligned}
|\psi_{\text{in}}\rangle=&(\alpha |1\rangle_1+\beta |0\rangle_1) \otimes (\delta |1\rangle_2+\gamma |0\rangle_2)\\&\otimes (\nu |1\rangle_3+\mu |0\rangle_3).
\end{aligned}
\end{eqnarray}
We first apply a spontaneous-parametric-down-conversion (SPDC) source to produce a polarization-entangled photon pair \cite{SPDC,SPDC1}.
After the pump pulse of the ultraviolet-light beam passes through two adjacent Type-I phase-matched $\beta$-barium borate (BBO) crystals, a two-photon pair  will occupy spatial modes $a$ and $d$ with the normalization state
\begin{eqnarray}\label{eq12}
|\psi_{\text{initial}}\rangle=(\alpha \hat a^{\dag}_{H}\hat d^{\dag}_{H}
+\beta \hat a^{\dag}_{V}\hat d^{\dag}_{V})|\textrm{vac}\rangle.
\end{eqnarray}
Next, $\textrm{BS}_{1}$ and $\textrm{BS}_{2}$ are set to yield  the transformations
\begin{eqnarray}\label{eq13}
\begin{aligned}
&\hat a^{\dag}_{H} \xrightarrow{\textrm{BS}_{1}} \gamma \hat a^{\dag}_{H}+\delta\hat b^{\dag}_{H},\;\;
 \hat a^{\dag}_{V} \xrightarrow{\textrm{BS}_{1}} \gamma \hat a^{\dag}_{V}+\delta\hat b^{\dag}_{V},\\
&\hat d^{\dag}_{H} \xrightarrow{\textrm{BS}_{2}} \mu \hat c^{\dag}_{H}+\nu\hat d^{\dag}_{H},\;\;
 \hat d^{\dag}_{V} \xrightarrow{\textrm{BS}_{2}} \mu \hat c^{\dag}_{V}+\nu\hat d^{\dag}_{V},
\end{aligned}
\end{eqnarray}
where $|\gamma|^{2}+|\delta|^{2}=|\mu|^{2}+|\nu|^{2}=1$.
Thus, $\textrm{BS}_{1}$ and $\textrm{BS}_{2}$ make $|\psi_{\text{initial}}\rangle$ change into
\begin{eqnarray}\label{eq14}
\begin{aligned}
|\psi_{\text{in}}\rangle=&(\alpha\gamma\mu \hat a^{\dag}_{H}\hat c^{\dag}_{H}
+\alpha\gamma\nu \hat a^{\dag}_{H}\hat d^{\dag}_{H}
+\alpha\delta\mu \hat b^{\dag}_{H}\hat c^{\dag}_{H}\\&
+\alpha\delta\nu \hat b^{\dag}_{H}\hat d^{\dag}_{H}
+\beta\gamma\mu \hat a^{\dag}_{V}\hat c^{\dag}_{V}
+\beta\gamma\nu \hat a^{\dag}_{V}\hat d^{\dag}_{V}\\&
+\beta\delta\mu \hat b^{\dag}_{V}\hat c^{\dag}_{V}
+\beta\delta\nu \hat b^{\dag}_{V}\hat d^{\dag}_{V})|\textrm{vac}\rangle.
\end{aligned}
\end{eqnarray}
This procedure is similar to the state preparation in Ref. \cite{path-polarization1}.
By defining our logical qubits as
\begin{eqnarray}\label{eq15}
\begin{aligned}
&|VV\rangle\equiv|0\rangle_{1},\;\;\; |HH\rangle\equiv|1\rangle_{1},\\
&|a\rangle\equiv|0\rangle_{2},\quad\;\;\;|b\rangle\equiv|1\rangle_{2},\\
&|c\rangle\equiv|0\rangle_{3},\quad\;\;\;|d\rangle\equiv|1\rangle_{3},
\end{aligned}
\end{eqnarray}
one can see that Eq. (\ref{eq14}) corresponds to the general input state described in  Eq. (\ref{eq11}).

On the basis the encoding described above, the controlled-SWAP operation can be performed with the apparatus shown in Fig. \ref{2b}.
In detail, $\textrm{PBS}_1$ and $\textrm{PBS}_2$ induce the following transformations:
\begin{eqnarray}\label{eq16}
\begin{aligned}
&\hat a^{\dag}_{V} \xrightarrow{\textrm{PBS}_1} \hat a^{\dag}_{V},\;\;
 \hat a^{\dag}_{H} \xrightarrow{\textrm{PBS}_1} \hat c^{\dag}_{H},\\
&\hat c^{\dag}_{V} \xrightarrow{\textrm{PBS}_1} \hat c^{\dag}_{V},\;\;\;
 \hat c^{\dag}_{H} \xrightarrow{\textrm{PBS}_1} \hat a^{\dag}_{H},\\
&\hat b^{\dag}_{V} \xrightarrow{\textrm{PBS}_2} \hat b^{\dag}_{V},\;\;\;
 \hat b^{\dag}_{H} \xrightarrow{\textrm{PBS}_2} \hat d^{\dag}_{H},\\
&\hat d^{\dag}_{V} \xrightarrow{\textrm{PBS}_2} \hat d^{\dag}_{V},\;\;
 \hat d^{\dag}_{H} \xrightarrow{\textrm{PBS}_2} \hat b^{\dag}_{H}.
\end{aligned}
\end{eqnarray}
Thus, PBS$_1$ and PBS$_2$ convert the state $|\psi_{\text{in}}\rangle$ described in Eq. (\ref{eq14}) into
\begin{eqnarray}\label{eq17}
\begin{aligned}
|\psi_{\text{out}}\rangle=&(\alpha\gamma\mu \hat a^{\dag}_{H}\hat c^{\dag}_{H}
+\alpha\gamma\nu \hat c^{\dag}_{H}\hat b^{\dag}_{H}
+\alpha\delta\mu \hat d^{\dag}_{H}\hat a^{\dag}_{H}\\&
+\alpha\delta\nu \hat d^{\dag}_{H}\hat b^{\dag}_{H}
+\beta\gamma\mu \hat a^{\dag}_{V}\hat c^{\dag}_{V}
+\beta\gamma\nu \hat a^{\dag}_{V}\hat d^{\dag}_{V}\\&
+\beta\delta\mu \hat b^{\dag}_{V}\hat c^{\dag}_{V}
+\beta\delta\nu \hat b^{\dag}_{V}\hat d^{\dag}_{V})|\textrm{vac}\rangle,
\end{aligned}
\end{eqnarray}
which corresponds to the output state
\begin{eqnarray}\label{eq18}
\begin{aligned}
|\psi_{\textrm{out}}\rangle=&(\alpha\gamma\mu |100\rangle
+\alpha\gamma\nu |110\rangle
+\alpha\delta\mu |101\rangle\\
&+\alpha\delta\nu |111\rangle
+\beta\gamma\mu |000\rangle
+\beta\gamma\nu |001\rangle\\
&+\beta\delta\mu |010\rangle
+\beta\delta\nu |011\rangle)_{123}.
\end{aligned}
\end{eqnarray}

Therefore, only two PBSs is sufficient to implement a Fredkin gate, which is much less than 14 linear optics in Ref. \cite{path-polarization1}.

\subsection{The generalized controlled-SWAP gate on $\mathbf{\mathbb{C}^2\otimes \mathbb{C}^3\otimes \mathbb{C}^3}$}\label{sec2.3}

We now extend the target system of controlled-SWAP gates to two three-dimensional Hilbert spaces $\mathbb{C}^3\otimes \mathbb{C}^3$.
The controlled-SWAP gate acting on the state in $\mathbb{C}^2\otimes \mathbb{C}^3\otimes \mathbb{C}^3$ can be described as
\begin{eqnarray}\label{eq19}
\begin{aligned}
U_{\text{CSWAP}}^{2,3,3}&=\sum_{i,j=0}^{2}(|0ij\rangle\langle0ij|+|1ij\rangle\langle1ji|),
\end{aligned}
\end{eqnarray}
where $i,j= 0,1,2$.

\begin{figure} [htbp]
  \centering
  \subfigure[State preparation]{
  \includegraphics[width=5.6 cm]{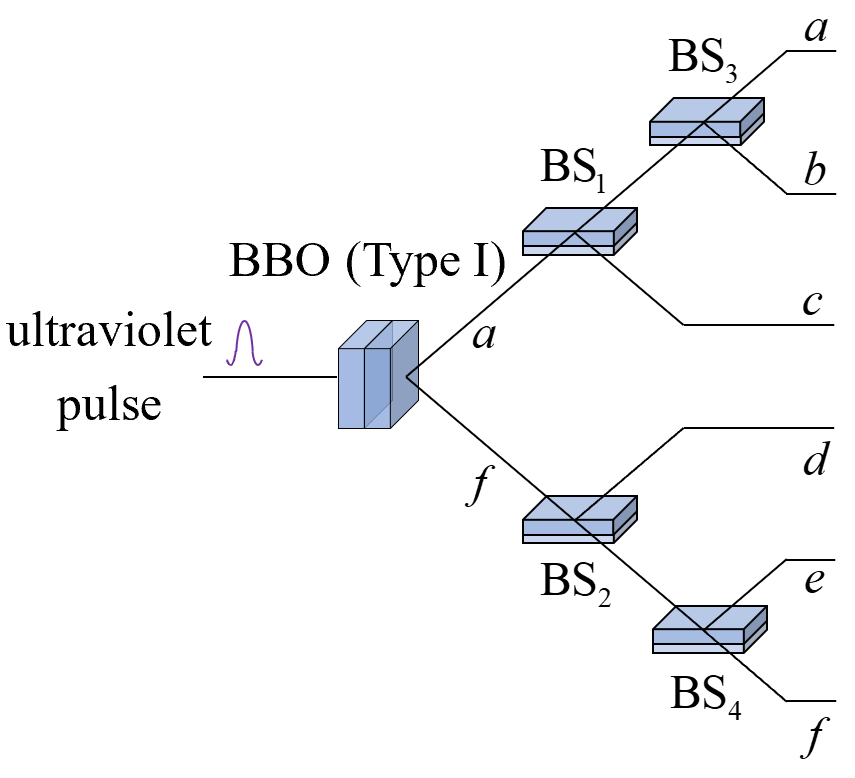}
  \label{3a}}
  \subfigure[Controlled-swap gate]{
  \includegraphics[width=4.0 cm]{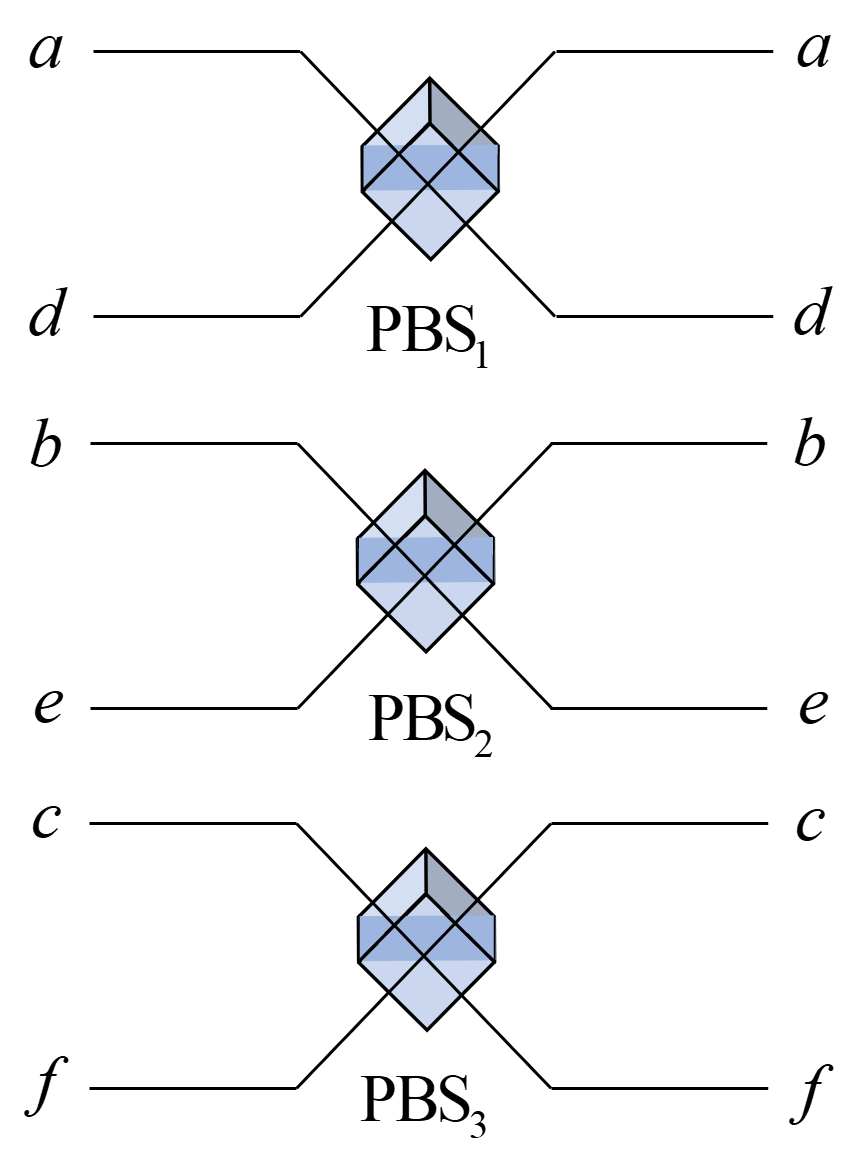}
  \label{3b}}
  \caption{(a) Experimental setup of the state preparation. (b) The linear-optical controlled-SWAP gate on $\mathbb{C}^2\otimes \mathbb{C}^3\otimes \mathbb{C}^3$.}
  \label{Fig.3}
\end{figure}

The general input state of the gate $U_{\text{CSWAP}}^{2,3,3}$ can be generated by means of Fig. \ref{3a}. As discussed in Sec. \ref{sec2.2}, a polarization-entangled photon pair $|\varphi_{\text{initial}}\rangle$ occuping spatial modes $a$ and $f$ can be generated by the SPDC source. Here
\begin{eqnarray}\label{eq20}
|\varphi_{\text{initial}}\rangle=(\alpha \hat a^{\dag}_{H}\hat f^{\dag}_{H}
+\beta \hat a^{\dag}_{V}\hat f^{\dag}_{V})|\textrm{vac}\rangle.
\end{eqnarray}
%

Subsequently, the two photons pass through the BS pairs ($\textrm{BS}_{1}$, $\textrm{BS}_{2}$) and ($\textrm{BS}_{3}$, $\textrm{BS}_{4}$) in succession. Here $\textrm{BS}_{1}$, $\textrm{BS}_{2}$, $\textrm{BS}_{3}$, and $\textrm{BS}_{4}$ complete the transformations
\begin{eqnarray}\label{eq21}
\begin{aligned}
&\hat a^{\dag}_{\Gamma} \xrightarrow{\textrm{BS}_{1}} x_{1} \hat a^{\dag}_{\Gamma} + y_{1} \hat c^{\dag}_{\Gamma},\;\;
\hat f^{\dag}_{\Gamma} \xrightarrow{\textrm{BS}_{2}} x_{2} \hat d^{\dag}_{\Gamma} + y_{2} \hat f^{\dag}_{\Gamma},\\
&\hat a^{\dag}_{\Gamma} \xrightarrow{\textrm{BS}_{3}} x_{3} \hat a^{\dag}_{\Gamma} + y_{3} \hat b^{\dag}_{\Gamma},\;\;
\hat f^{\dag}_{\Gamma} \xrightarrow{\textrm{BS}_{4}} x_{4} \hat e^{\dag}_{\Gamma} + y_{4} \hat f^{\dag}_{\Gamma},
\end{aligned}
\end{eqnarray}
where $\Gamma\in\{H,V\}$, and $|x_{k}|^{2}+|y_{k}|^{2}=1$ $(k=1,2,3,4)$.
Hence, the operations transform the state $|\varphi_{\text{initial}}\rangle$ into
\begin{eqnarray}\label{eq22}
\begin{aligned}
|\varphi_{\text{in}}\rangle=&
(\theta_{100}  \hat a^{\dag}_{H}\hat d^{\dag}_{H}
+\theta_{101}  \hat a^{\dag}_{H}\hat e^{\dag}_{H}
+\theta_{102}  \hat a^{\dag}_{H}\hat f^{\dag}_{H}\\&
+\theta_{110}  \hat b^{\dag}_{H}\hat d^{\dag}_{H}
+\theta_{111}  \hat b^{\dag}_{H}\hat e^{\dag}_{H}
+\theta_{112}  \hat b^{\dag}_{H}\hat f^{\dag}_{H}\\&
+\theta_{120}  \hat c^{\dag}_{H}\hat d^{\dag}_{H}
+\theta_{121}  \hat c^{\dag}_{H}\hat e^{\dag}_{H}
+\theta_{122}  \hat c^{\dag}_{H}\hat f^{\dag}_{H}\\&
+\theta_{000} \hat a^{\dag}_{V}\hat d^{\dag}_{V}
+\theta_{001} \hat a^{\dag}_{V}\hat e^{\dag}_{V}
+\theta_{002} \hat a^{\dag}_{V}\hat f^{\dag}_{V}\\&
+\theta_{010} \hat b^{\dag}_{V}\hat d^{\dag}_{V}
+\theta_{011} \hat b^{\dag}_{V}\hat e^{\dag}_{V}
+\theta_{012} \hat b^{\dag}_{V}\hat f^{\dag}_{V}\\&
+\theta_{020} \hat c^{\dag}_{V}\hat d^{\dag}_{V}
+\theta_{021} \hat c^{\dag}_{V}\hat e^{\dag}_{V}
+\theta_{022} \hat c^{\dag}_{V}\hat f^{\dag}_{V})|\textrm{vac}\rangle.
\end{aligned}
\end{eqnarray}
with
\begin{eqnarray}\label{eq23}
\begin{aligned}
&\theta_{100}=\alpha x_{1} x_{2} x_{3},\;\;\;\;\;\;\;
\theta_{101}=\alpha x_{1} y_{2} x_{3} x_{4},\\&
\theta_{102}=\alpha x_{1} y_{2} x_{3} y_{4},\;\;\;\;
\theta_{110}=\alpha x_{1} x_{2} y_{3},\\&
\theta_{111}=\alpha x_{1} y_{2} y_{3} x_{4},\;\;\;\;
\theta_{112}=\alpha x_{1} y_{2} y_{3} y_{4},\\&
\theta_{120}=\alpha y_{1} x_{2},\;\;\;\;\;\;\;\;\;\;\;
\theta_{121}=\alpha y_{1} y_{2} x_{4},\\&
\theta_{122}=\alpha y_{1} y_{2} y_{4},\;\;\;\;\;\;\;\;
\theta_{000}=\beta x_{1} x_{2} x_{3},\\&
\theta_{001}=\beta x_{1} y_{2} x_{3} x_{4},\;\;\;\;
\theta_{002}=\beta x_{1} y_{2} x_{3} y_{4},\\&
\theta_{010}=\beta x_{1} x_{2} y_{3},\quad\;\;\;\;
\theta_{011}=\beta x_{1} y_{2} y_{3} x_{4},\\&
\theta_{012}=\beta x_{1} y_{2} y_{3} y_{4},\;\;\;\;\;
\theta_{020}=\beta y_{1} x_{2},\\&
\theta_{021}=\beta y_{1} y_{2} x_{4},\;\;\;\;\;\;\;\;
\theta_{022}=\beta y_{1} y_{2} y_{4}.
\end{aligned}
\end{eqnarray}
On the basis of the encoding rules
\begin{eqnarray}\label{eq24}
\begin{aligned}
&
|VV\rangle\equiv|0\rangle_{1}, |HH\rangle\equiv|1\rangle_{1},\\&
|a\rangle\equiv|0\rangle_{2},\;\;\;\; |b\rangle\equiv|1\rangle_{2},\;\;\;\;\; |c\rangle\equiv|2\rangle_{2},\\&
|d\rangle\equiv|0\rangle_{3},\;\;\;\; |e\rangle\equiv|1\rangle_{3},\;\;\;\;\; |f\rangle\equiv|2\rangle_{3},
\end{aligned}
\end{eqnarray}
one can see that Eq. (\ref{eq22}) corresponds to the general input logical state
\begin{eqnarray}\label{eq25}
\begin{aligned}
|\varphi_{\textrm{in}}\rangle=&
(\theta_{100} |100\rangle
+\theta_{101} |101\rangle
+\theta_{102} |102\rangle\\&
+\theta_{110} |110\rangle
+\theta_{111} |111\rangle
+\theta_{112} |112\rangle\\&
+\theta_{120} |120\rangle
+\theta_{121} |121\rangle
+\theta_{122} |122\rangle\\&
+\theta_{000} |000\rangle
+\theta_{001} |001\rangle
+\theta_{002} |002\rangle\\&
+\theta_{010} |010\rangle
+\theta_{011} |011\rangle
+\theta_{012} |012\rangle\\&
+\theta_{020} |020\rangle
+\theta_{021} |021\rangle
+\theta_{022} |022\rangle)_{123}.
\end{aligned}
\end{eqnarray}
We note that Eq. (\ref{eq25}) is just the expanded version of the general input
state on $\mathbb{C}^2\otimes \mathbb{C}^2\otimes \mathbb{C}^2$ described in Eq.(\ref{eq11}).

On the basis of Eq. (\ref{eq22}) and Eq. (\ref{eq24}), $U_{\text{CSWAP}}^{2,3,3}$ can be efficiently constructed by use of three PBSs as shown in Fig. \ref{3b}, where the PBSs result in
\begin{eqnarray}\label{eq26}
\begin{aligned}
&\hat a^{\dag}_{V} \xrightarrow{\textrm{PBS}_1} \hat a^{\dag}_{V},\;\;\;\;
 \hat a^{\dag}_{H} \xrightarrow{\textrm{PBS}_1} \hat d^{\dag}_{H},\\
&\hat d^{\dag}_{V} \xrightarrow{\textrm{PBS}_1} \hat d^{\dag}_{V},\;\;\;\;
 \hat d^{\dag}_{H} \xrightarrow{\textrm{PBS}_1} \hat a^{\dag}_{H},\\
&\hat b^{\dag}_{V} \xrightarrow{\textrm{PBS}_2} \hat b^{\dag}_{V},\;\;\;\;\;
 \hat b^{\dag}_{H} \xrightarrow{\textrm{PBS}_2} \hat e^{\dag}_{H},\\
&\hat e^{\dag}_{V} \xrightarrow{\textrm{PBS}_2} \hat e^{\dag}_{V},\;\;\;\;
 \hat e^{\dag}_{H} \xrightarrow{\textrm{PBS}_2} \hat b^{\dag}_{H},\\
&\hat c^{\dag}_{V} \xrightarrow{\textrm{PBS}_3} \hat c^{\dag}_{V},\;\;\;\;
 \hat c^{\dag}_{H} \xrightarrow{\textrm{PBS}_3} \hat f^{\dag}_{H},\\
&\hat f^{\dag}_{V} \xrightarrow{\textrm{PBS}_3} \hat f^{\dag}_{V},\;\;\;\;
 \hat f^{\dag}_{H} \xrightarrow{\textrm{PBS}_3} \hat c^{\dag}_{H}.
\end{aligned}
\end{eqnarray}
After the wavepackets have been mixed at PBS$_1$, PBS$_2$, and PBS$_3$,  the initial state $|\varphi_{\text{in}}\rangle$  will be
\begin{eqnarray}\label{eq27}
\begin{aligned}
|\varphi_{\text{out}}\rangle=\;&
(\theta_{100}  \hat a^{\dag}_{H}\hat d^{\dag}_{H}
+\theta_{101}  \hat b^{\dag}_{H}\hat d^{\dag}_{H}
+\theta_{102}  \hat c^{\dag}_{H}\hat d^{\dag}_{H}\\&
+\theta_{110}  \hat a^{\dag}_{H}\hat e^{\dag}_{H}
+\theta_{111}  \hat b^{\dag}_{H}\hat e^{\dag}_{H}
+\theta_{112}  \hat c^{\dag}_{H}\hat e^{\dag}_{H}\\&
+\theta_{120}  \hat a^{\dag}_{H}\hat f^{\dag}_{H}
+\theta_{121}  \hat b^{\dag}_{H}\hat f^{\dag}_{H}
+\theta_{122}  \hat c^{\dag}_{H}\hat f^{\dag}_{H}\\&
+\theta_{000} \hat a^{\dag}_{V}\hat d^{\dag}_{V}
+\theta_{001} \hat a^{\dag}_{V}\hat e^{\dag}_{V}
+\theta_{002} \hat a^{\dag}_{V}\hat f^{\dag}_{V}\\&
+\theta_{010} \hat b^{\dag}_{V}\hat d^{\dag}_{V}
+\theta_{011} \hat b^{\dag}_{V}\hat e^{\dag}_{V}
+\theta_{012} \hat b^{\dag}_{V}\hat f^{\dag}_{V}\\&
+\theta_{020} \hat c^{\dag}_{V}\hat d^{\dag}_{V}
+\theta_{021} \hat c^{\dag}_{V}\hat e^{\dag}_{V}
+\theta_{022} \hat c^{\dag}_{V}\hat f^{\dag}_{V})|\textrm{vac}\rangle,
\end{aligned}
\end{eqnarray}
which corresponds to the output state
\begin{eqnarray}\label{eq28}
\begin{aligned}
|\varphi_{\textrm{out}}\rangle=\;&
(\theta_{100} |100\rangle
+\theta_{101} |110\rangle
+\theta_{102} |120\rangle\\&
+\theta_{110} |101\rangle
+\theta_{111} |111\rangle
+\theta_{112} |121\rangle\\&
+\theta_{120} |102\rangle
+\theta_{121} |112\rangle
+\theta_{122} |122\rangle\\&
+\theta_{000} |000\rangle
+\theta_{001} |001\rangle
+\theta_{002} |002\rangle\\&
+\theta_{010} |010\rangle
+\theta_{011} |011\rangle
+\theta_{012} |012\rangle\\&
+\theta_{020} |020\rangle
+\theta_{021} |021\rangle
+\theta_{022} |022\rangle)_{123}.
\end{aligned}
\end{eqnarray}

\subsection{The generalized controlled-SWAP gate on $\mathbf{\mathbb{C}^2\otimes \mathbb{C}^{\emph{d}}\otimes \mathbb{C}^{\emph{d}}}$}\label{sec2.4}

The device in Fig. \ref{Fig.3}  can be extended to implement the controlled-SWAP gate $U_{\text{CSWAP}}^{2,d,d}$ on  $\mathbb{C}^2\otimes \mathbb{C}^d\otimes \mathbb{C}^d$, where
\begin{eqnarray}\label{eq29}
\begin{aligned}
U_{\text{CSWAP}}^{2,d,d}=\sum_{i,j=0}^{d-1}(|0ij\rangle\langle0ij|+|1ij\rangle\langle1ji|),
\end{aligned}
\end{eqnarray}
where $i,j=0,1,2,\cdots,d-1$.

\begin{figure} [htbp]
  \centering
  \subfigure[State preparation]{
  \includegraphics[width=5.6 cm]{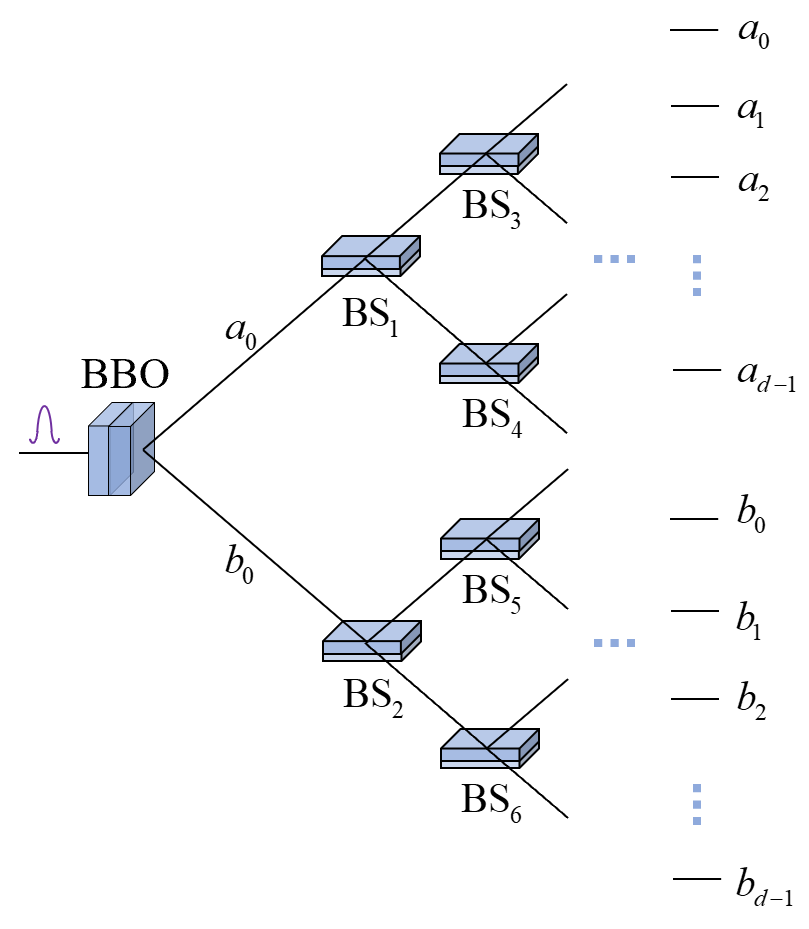}
  \label{4a}}
  \subfigure[Controlled-swap gate]{
  \includegraphics[width=4 cm]{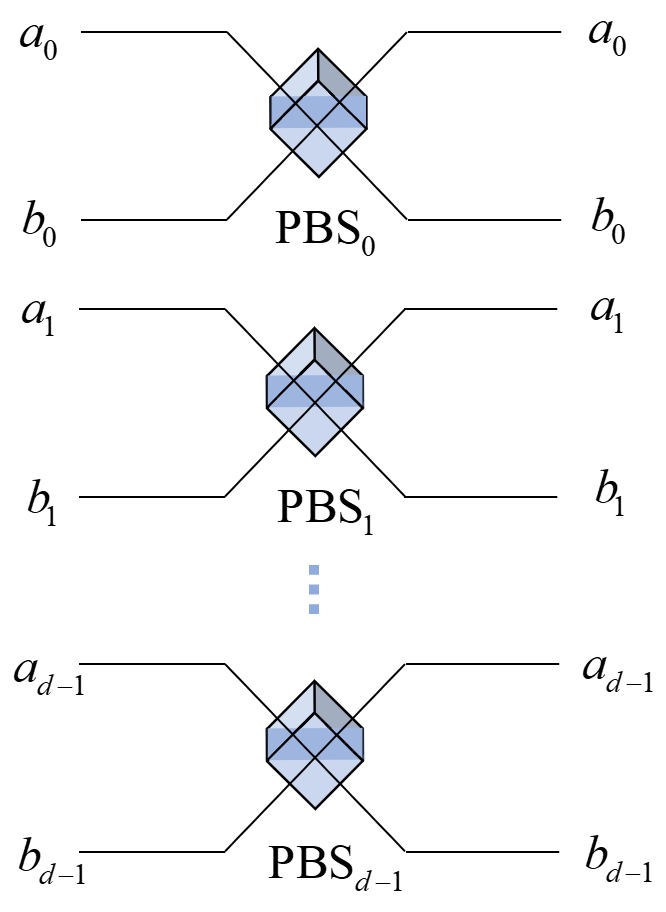}
  \label{4b}}
  \caption{(a) Experimental setup of the state-preparation. (b) The controlled-SWAP gate on $\mathbb{C}^2\otimes \mathbb{C}^d\otimes \mathbb{C}^d$.}
  \label{Fig.4}
\end{figure}

A similar argument as made in Sec. \ref{sec2.3}, Fig. \ref{4a} presents the expanded version of the input state. We suppose that $d=2^{n}+q$, where $0\leq q < 2^{n}$ and $n$ is a positive integer. The SPDC source produces a polarization-entangled state
\begin{eqnarray}\label{eq30}
|\omega_{\text{initial}}\rangle=(\alpha \hat a^{\dag}_{0H}\hat b^{\dag}_{0H}+\beta \hat a^{\dag}_{0V}\hat b^{\dag}_{0V})|\textrm{vac}\rangle,
\end{eqnarray}
in the spatial modes $a_{0}$ and $b_{0}$.

Subsequently, the photon in the spatial path $a_{0}$ ($b_{0}$) is  driven through  $\textrm{BS}_1$ ($\textrm{BS}_2$), yielding  spatial basis states $\{|a_{0}\rangle, |a_{1}\rangle\}$ ($\{|b_{0}\rangle,|b_{1}\rangle\}$).
Next, $\textrm{BS}_3$ and $\textrm{BS}_4$ ($\textrm{BS}_5$ and $\textrm{BS}_6$) are set in paths $|a_{0}\rangle$ and $|a_{1}\rangle$ ($|b_{0}\rangle$ and $|b_{1}\rangle$) to generate four spatial basis states $\{|a_{0}\rangle, |a_{1}\rangle, |a_{2}\rangle, |a_{3}\rangle\}$ ($\{|b_{0}\rangle, |b_{1}\rangle, |b_{2}\rangle, |b_{3}\rangle\}$).
The triangular array of BSs is expanded until the $2^{n}$ spatial basis states $\{|a_{0}\rangle,|a_{1}\rangle,\cdots,|a_{2^{n}-1}\rangle\}$
($\{|b_{0}\rangle,|b_{1}\rangle,\cdots,|b_{2^{n}-1}\rangle\}$) are obtained.

Finally, $q$ $\textrm{BS}$s are set to $q$ spatial modes of $\{a_{0}, a_{1}, \cdots, a_{2^{n}-1}\}$ separately to produce $d$ spatial basis states $\{|a_{0}\rangle,|a_{1}\rangle,\cdots,|a_{d-1}\rangle\}$, and for the same is done for $\{b_{0}, b_{1}, \cdots, b_{2^{n}-1}\}$.
Note that this step is not necessary if $q=0$.
Putting all the pieces together, one can see that $2(d-1)$ $\textrm{BS}$s are applied, and the state $|\omega_{\text{initial}}\rangle$ becomes
\begin{eqnarray}\label{eq31}
\begin{aligned}
|\omega_{\text{in}}\rangle=\sum_{i,j=0}^{d-1}
(\theta_{1ij} \hat a^{\dag}_{iH} \hat b^{\dag}_{jH}
+\theta_{0ij} \hat a^{\dag}_{iV} \hat b^{\dag}_{jV})|\textrm{vac}\rangle.
\end{aligned}
\end{eqnarray}
Here $\sum_{i,j=0}^{d-1}(|\theta_{1ij}|^2+|\theta_{0ij}|^2)=1$.  The complex parameters $\theta_{0ij}$ and  $\theta_{1ij}$ are determined by $\alpha$, $\beta$, and reflection and transmission parameters of the $2(d-1)$ $\textrm{BS}$s.

If we encode the computing logic qubits as
\begin{eqnarray}\label{eq32}
\begin{aligned}
&|VV\rangle\equiv|0\rangle_{1},    |HH\rangle\equiv|1\rangle_{1},\\&
|a_{i}\rangle\equiv|i\rangle_{2}, \;\;\; |b_{j}\rangle\equiv|j\rangle_{3} (i,j=0,1,\cdots,d-1).
\end{aligned}
\end{eqnarray}
The input state of $U_{\text{CSWAP}}^{2,d,d}$ described in Eq. (\ref{eq31}) corresponds to
\begin{eqnarray}\label{eq33}
\begin{aligned}
|\omega_{\textrm{in}}\rangle=\sum_{i,j=0}^{d-1}(\theta_{1ij} |1ij\rangle+\theta_{0ij}|0ij\rangle).
\end{aligned}
\end{eqnarray}
As shown in Fig. \ref{4b}, $U_{\text{CSWAP}}^{2,d,d}$ can be implemented by our using solely $d$ PBSs.
Specifically, when the control logic qubit is in the state $|0\rangle_{1}$ corresponding to $|VV\rangle$, the spatial-based target states will not be changed by the PBSs, i.e.,
\begin{eqnarray}\label{eq34}
|0ij\rangle_{123} \xrightarrow{\textrm{PBSs}} |0ij\rangle_{123}\;(i,j=0,1,\cdots,d-1).
\end{eqnarray}
When the control logic qubit is in the state $|1\rangle_{1}$ corresponding to $|HH\rangle$, the target states will be exchanged by the PBSs, i.e.,
\begin{eqnarray}\label{eq35}
\begin{aligned}
\hat a^{\dag}_{iH}\hat b^{\dag}_{jH} \xrightarrow{\textrm{PBSs}} \hat a^{\dag}_{jH}\hat b^{\dag}_{iH}\;(i,j=0,1,\cdots,d-1),
\end{aligned}
\end{eqnarray}
which corresponds to
\begin{eqnarray}\label{eq36}
|1ij\rangle_{123} \xrightarrow{\textrm{PBSs}} |1ji\rangle_{123}\;(i,j=0,1,\cdots,d-1).
\end{eqnarray}
On the basis of Eqs. (\ref{eq34}) and (\ref{eq36}), it can be deduced that
\begin{eqnarray}\label{eq37}
\begin{aligned}
|\omega_{\textrm{in}}\rangle\xrightarrow{\textrm{PBSs}}|\omega_{\textrm{out}}\rangle=\sum_{i,j=0}^{d-1}(\theta_{1ij}|1ji\rangle+\theta_{0ij}|0ij\rangle).
\end{aligned}
\end{eqnarray}
Therefore, only $d$ PBSs can efficiently complete the operation $U_{\text{CSWAP}}^{2,d,d}$.

\section{Average fidelity of the controlled-SWAP gate}\label{sec3}

The schemes presented all are constructed by use of solely PBSs, and these PBSs can be replaced with calcite beam displacers (BDs) as they all essentially direct the photons into different spatial modes on the basis of the polarization states. To conveniently compare the performance in Fig. \ref{2b} with that of the scheme proposed in Ref. \cite{path-polarization1}, we consider the imperfections induced by the PBS ($\bar U^{\dag}_{\textrm{PBS}}$) to be the same as those induced by the calcite beam displacer ($\bar U^{\dag}_{\textrm{BD}}$).
That is, $\bar U^{\dag}_{\textrm{PBS}}$ accomplishes the following transformations \cite{path-polarization1}:
\begin{eqnarray}\label{eq38}
\begin{aligned}
\hat \chi^{\dag}_{H}  \xrightarrow{\bar U^{\dag}_{\textrm{PBS}}}&
\frac{1}{\sqrt{1+|r|}}[(\textrm{cos}\theta-\sqrt{r}\textrm{sin}\theta) \hat \eta^{\dag}_{H} \\&
-(\textrm{sin}\theta+\sqrt{r}^{*}\textrm{cos}\theta) \hat \eta^{\dag}_{V}],
\end{aligned}
\end{eqnarray}
\begin{eqnarray}\label{eq39}
\begin{aligned}
\hat \eta^{\dag}_{H}  \xrightarrow{\bar U^{\dag}_{\textrm{PBS}}}&
\frac{1}{\sqrt{1+|r|}}[(\textrm{cos}\theta-\sqrt{r}\textrm{sin}\theta) \hat \chi^{\dag}_{H} \\&
-(\textrm{sin}\theta+\sqrt{r}^{*}\textrm{cos}\theta) \hat \chi^{\dag}_{V}],
\end{aligned}
\end{eqnarray}
\begin{eqnarray}\label{eq40}
\begin{aligned}
\hat \chi^{\dag}_{V}  \xrightarrow{\bar U^{\dag}_{\textrm{PBS}}}&
\frac{1}{\sqrt{1+|r|}}[(\textrm{sin}\theta+\sqrt{r}\textrm{cos}\theta) \hat \chi^{\dag}_{H} \\&
+(\textrm{cos}\theta-\sqrt{r}^{*}\textrm{sin}\theta) \hat \chi^{\dag}_{V}],
\end{aligned}
\end{eqnarray}
\begin{eqnarray}\label{eq41}
\begin{aligned}
\hat \eta^{\dag}_{V}  \xrightarrow{\bar U^{\dag}_{\textrm{PBS}}}&
\frac{1}{\sqrt{1+|r|}}[(\textrm{sin}\theta+\sqrt{r}\textrm{cos}\theta) \hat \eta^{\dag}_{H} \\&
+(\textrm{cos}\theta-\sqrt{r}^{*}\textrm{sin}\theta) \hat \eta^{\dag}_{V}],
\end{aligned}
\end{eqnarray}
where $\theta$ and $r$ are the deviation of mirror mounts and the polarization extinction ratio of the PBS, respectively,
$(\chi,\eta)=(a,c)$ or $(\chi,\eta)=(b,d)$, and $\sqrt{r}^{*}$ denotes the conjugate complex of $\sqrt{r}$.

\begin{figure} [htbp]
  \centering
  \includegraphics[width=8cm]{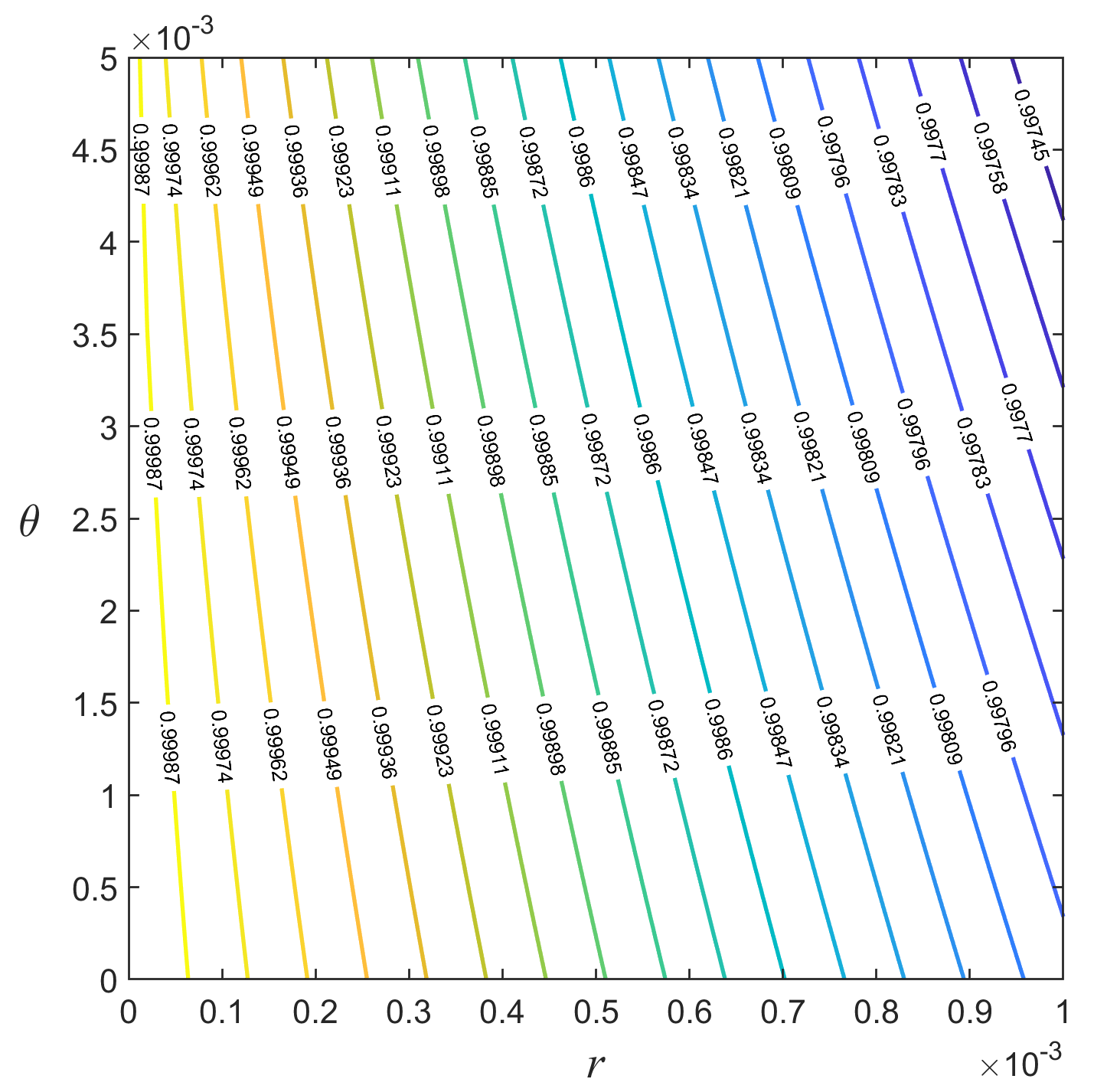}
  \caption{Average fidelity $\overline{F}$ of the controlled-SWAP gate $U_{\text{CSWAP}}^{2,2,2}$ with $r\in[0,1\times10^{-3}]$ and $\theta\in[0,5\times10^{-3}]$.}
  \label{Fig.5}
\end{figure}

The performance of the presented controlled-SWAP gate $U_{\text{CSWAP}}^{2,2,2}$ can be evaluated by 
\begin{eqnarray}\label{eq42}
\begin{aligned}
\overline{F}&=\frac{1}{(2\pi)^{3}}\int_0^{2\pi}\int_0^{2\pi}\int_0^{2\pi}
|\langle\psi_{\textrm{real}}|\psi_{\textrm{out}}\rangle|^{2}
\textrm{d}x\textrm{d}y\textrm{d}z\\
&=\frac{(\textrm{cos}\theta
-\sqrt{r}\textrm{sin}\theta)^{4}}{(1+|r|)^{2}}
+\frac{3}{16}\cdot\frac{(\textrm{sin}\theta
+\sqrt{r}\textrm{cos}\theta)^{4}}{(1+|r|)^{2}},
\end{aligned}
\end{eqnarray}
where the ideal output state $|\psi_{\textrm{out}}\rangle$ is given by Eq. (\ref{eq18}) and $|\psi_{\textrm{real}}\rangle$ represents the practical
output state with $\alpha=\textrm{cos}x$, $\beta=\textrm{sin}x$, $\gamma=\textrm{cos}y$,
$\delta=\textrm{sin}y$, $\mu=\textrm{cos}z$, and $\upsilon=\textrm{sin}z$.
Figure \ref{Fig.5} shows the average fidelity $\overline{F}$ of $U_{\text{CSWAP}}^{2,2,2}$ as a function of $r$ and $\theta$.
In particular, when $r=0.001$ and $\theta=0.005\textrm{rad}$, the average fidelity $\overline{F}=0.997$.

For the input state $|\psi^{\prime}\rangle=|000\rangle$ of $U_{\text{CSWAP}}^{2,2,2}$ corresponding to $\hat a^{\dag}_{V}\hat c^{\dag}_{V}|\textrm{vac}\rangle$, the ideal output state $|\psi^{\prime}_{\textrm{ideal}}\rangle $ and the practical $|\psi^{\prime}_{\textrm{real}}\rangle$ can be calculated as
\begin{eqnarray}\label{eq43}
\begin{aligned}
|\psi^{\prime}_{\textrm{ideal}}\rangle=\hat a^{\dag}_{V}\hat c^{\dag}_{V}|\textrm{vac}\rangle,
\end{aligned}
\end{eqnarray}
\begin{eqnarray}\label{eq44}
|\psi^{\prime}_{\textrm{real}}\rangle
&=&\frac{1}{1+|r|}
[(\textrm{sin}\theta+\sqrt{r}\textrm{cos}\theta)^{2}
\hat a^{\dag}_{H} \hat c^{\dag}_{H}\nonumber \\&&
+(\textrm{sin}\theta+\sqrt{r}\textrm{cos}\theta)
(\textrm{cos}\theta-\sqrt{r}^{*}\textrm{sin}\theta)
\hat a^{\dag}_{H} \hat c^{\dag}_{V}\nonumber \\&&
+(\textrm{cos}\theta-\sqrt{r}^{*}\textrm{sin}\theta)
(\textrm{sin}\theta+\sqrt{r}\textrm{cos}\theta)
\hat a^{\dag}_{V} \hat c^{\dag}_{H}\nonumber \\&&
+(\textrm{cos}\theta-\sqrt{r}^{*}\textrm{sin}\theta)^{2}
\hat a^{\dag}_{V} \hat c^{\dag}_{V}]|\textrm{vac}\rangle.
\end{eqnarray}
Thus, the average fidelity of $U_{\text{CSWAP}}^{2,2,2}$ for the input state $|000\rangle$ is given by
\begin{eqnarray}\label{eq45}
\overline{F}_{000}=|\langle\psi^{\prime}_{\textrm{real}}|\psi^{\prime}_{\textrm{ideal}}\rangle|^{2}
=\bigg|\frac{(\textrm{cos}\theta-\sqrt{r}\textrm{sin}\theta)^{2}}{1+|r|}\bigg|^{2}.
\end{eqnarray}
It is easy  to verify that $\overline{F}_{000}=\overline{F}_{001}=\overline{F}_{010}=\overline{F}_{011}=
\overline{F}_{100}=\overline{F}_{101}=\overline{F}_{110}=\overline{F}_{111}$.

\begin{figure} [htbp]
  \centering
  \subfigure[]{
  \includegraphics[width=6.35cm]{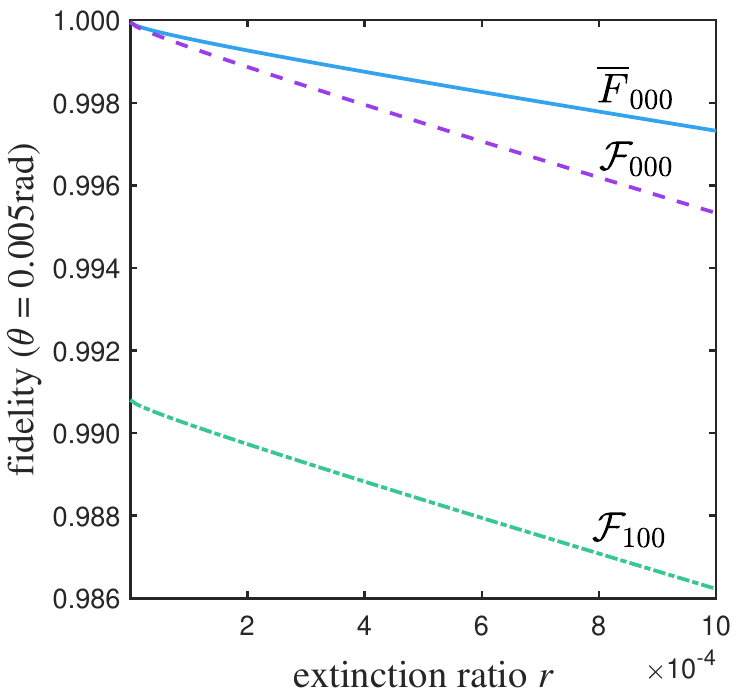}
  \label{6a}}
  \subfigure[]{
  \includegraphics[width=6.35cm]{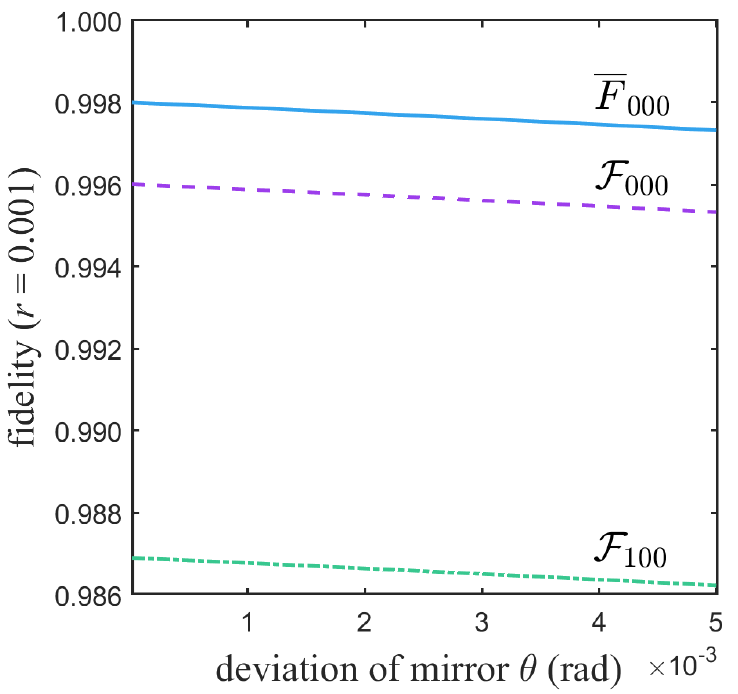}
  \label{6b}}
  \caption{(a) Fidelities of controlled-SWAP gate $U_{\text{CSWAP}}^{2,2,2}$ as a function of the extinction ratio $r$ with $\theta=5\times10^{-3}\textrm{rad}$.
  (b) Fidelities of controlled-SWAP gate $U_{\text{CSWAP}}^{2,2,2}$ as a function of the deviation of mirror mounts $\theta$ with $r=1\times10^{-3}$.
  The solid blue curves are for our $\overline{F}_{0ij}$ or $\overline{F}_{1ij}$, the dashed purple curves are for $\mathcal{F}_{0ij}$ \cite{path-polarization1}, and the dash-dotted green curves are for $\mathcal{F}_{1ij}$ \cite{path-polarization1}, where $i,j\in(0,1)$.}
  \label{Fig.6}
\end{figure}

Figure \ref{Fig.6} shows that $\overline{F}_{000}$ is much greater than $\mathcal{F}_{000}$, $\mathcal{F}_{001}$, $\mathcal{F}_{010}$, $\mathcal{F}_{011}$, $\mathcal{F}_{100}$, $\mathcal{F}_{101}$, $\mathcal{F}_{110}$, and $\mathcal{F}_{111}$ presented in Ref. \cite{path-polarization1}.
Moreover,
\begin{eqnarray}\label{eq46}
\begin{aligned}
\mathcal{F}_{000}&=\mathcal{F}_{001}=\mathcal{F}_{010}=\mathcal{F}_{011}\\&=
\bigg|\frac{(\textrm{cos}\theta -\sqrt{r}\textrm{sin}\theta)^{2}}{1+|r|}\bigg|^{2} \times \bigg|\frac{1}{1+r}\bigg|^{2},
\end{aligned}
\end{eqnarray}
\begin{eqnarray}\label{eq47}
\begin{aligned}
\mathcal{F}_{100}&=\mathcal{F}_{101}=\mathcal{F}_{110}=\mathcal{F}_{111}\\&=
\mathcal{F}_{000}\times\Bigg|\bigg(\frac{\textrm{i}(1+\epsilon)
(1-e^{\textrm{i}(\pi-\Delta\phi)})}
{2+2\epsilon+\epsilon^2}\bigg)^4\Bigg|^2,
\end{aligned}
\end{eqnarray}
where the imperfect parameters $\epsilon$ and $\Delta\phi$ are induced by BSs and phase shifters (PSs), respectively.
Although the presented $\overline{F}_{000}$ is slightly greater than $\mathcal{F}_{000}$ and $\mathcal{F}_{100}$ as shown in Fig. \ref{Fig.6} (where $\epsilon=0.02,\Delta\phi=\pi/36$, see Ref. \cite{path-polarization1}), $\overline{F}_{000}$ is affected only by $r$ and $\theta$, whereas errors $\epsilon$ and $\Delta\phi$ are eliminated.

\section{Discussion and Summary}\label{sec4}

Quantum logic gates are at the heart of quantum computing, and it has recently been demonstrated that use of multi-DOF encoding to implement quantum logic gates not only reduces the number of steps required to manipulate quantum states but also increases the fidelity of the logic gate \cite{path-polarization1,path-polarization2}.
In this work,  we design deterministic schemes for implementing a CNOT gate and a family of controlled-SWAP gates based on linear optics. 
The proposed schemes have several advantages over the synthesis-based one in Ref. \cite{Wei2020}, such as fewer linear optical elements and photonic resources, lower depth, and higher fidelity.
The control qubit and the target qubit (qudits) of the CNOT (controlled-SWAP) gate are encoded in the polarization and spatial states of a single photon (an entangled photon pair), respectively.

\begin{table}[htbp]
\centering
\caption{Comparison of CNOT and controlled-SWAP gates with the gates reported in Ref. \cite{path-polarization1}. }\label{Table1}
\begin{tabular}{ccccc}
\hline  \hline
  &  \makebox[0.13\textwidth]{\multirow{2}{*}{Study}} & \makebox[0.13\textwidth]{\multirow{2}{*}{Linear optics}}  & \makebox[0.1\textwidth]{Number of}   & \makebox[0.08\textwidth]{\multirow{2}{*}{Depth}}  \\
  &    &   &  elements  &   \\
\hline
\multirow{2}{*}{$U_{\text{CNOT}}^{2,2}$}     & Meng \cite{path-polarization1} & BD,BS,PS & 5                 & 5 \\ 
                                             & This work                      & PBS      & \textbf{1}        & \textbf{1} \\
 \cline{2-5}
\multirow{2}{*}{$U_{\text{CSWAP}}^{2,2,2}$} & Meng \cite{path-polarization1} & BD,BS,PS & 14                & 11 \\ 
                                             & This work                      & PBS      & \textbf{2}        & \textbf{1} \\
 \cline{2-5}
\multirow{2}{*}{$U_{\text{CSWAP}}^{2,d,d}$} & Meng \cite{path-polarization1} & none     & none              & none\\ 
                                             & This work                      & PBS      & \textbf{\emph{d}} & \textbf{1}\\ 
\hline  \hline
\end{tabular}
\end{table}

As shown in Table \ref{Table1}, the presented CNOT gate $U_{\text{CNOT}}^{2,2}$ can be completed by use of one PBS, and the depth of the circuit is 1. The same work reported in Ref. \cite{path-polarization1} requires two BDs, two BSs, and one PS, and the optical depth is 5.
Our three-qubit controlled-SWAP gate $U_{\text{CSWAP}}^{2,2,2}$ can be implemented with 2 PBSs, beating the earlier scheme of 14 linear optics \cite{path-polarization1}.
The optical depth of 1 for the gate $U_{\text{CSWAP}}^{2,2,2}$  is much lower than 11 in Ref. \cite{path-polarization1}.
Then we extend the program to $U_{\text{CSWAP}}^{2,d,d}$. Remarkably, the extended architecture requires $d$ PBSs and the $d$-independent optical depth is 1.
Moreover, the fidelity of $U_{\text{CSWAP}}^{2,2,2}$ shown in Fig. 6 is more than 99.7\%, which is also slightly higher than 99\% in Ref. \cite{path-polarization1},
and the error can be further reduced by the quantum Zeno effect \cite{QZE}.

The presented schemes for linear optical control gates are effective and easy to achieve with current technology.
With distinct merits of multi-DOF encoding, our effective hybrid schemes have strong potential for high-dimensional quantum computing and quantum communication.

\medskip

\section*{ACKNOWLEDGEMENTS} \par

This work was supported by the National Natural Science Foundation of China under Grant No. 62371038 and the Fundamental Research Funds for the Central Universities under Grant No. FRF-TP-19-011A3.

\end{document}